\documentclass{article}
\usepackage{spconf}
\usepackage{flushend}  
\usepackage{blindtext, graphicx}
\usepackage{bm}
\usepackage{booktabs}
\usepackage{subfig}
\usepackage[font=footnotesize]{caption}
\usepackage{hyperref}
\usepackage{xspace}
\usepackage[binary-units]{siunitx}
\usepackage{nicefrac}
\usepackage{tikz}

\usepackage[style=ieee,maxbibnames=2,minbibnames=1]{biblatex}
\AtEveryBibitem{%
\ifentrytype{inproceedings}{
    \clearfield{address}%
}{}
}

\usetikzlibrary{arrows}

\usepackage{localmath}

\newcommand{\smin}{\sigma_{\min}}
\newcommand{\smax}{\sigma_{\max}}
\newcommand{\mAhat}{\wh{\mA}}

\providecommand{\vsbar}{\bar{\vs}}
\providecommand{\mPa}{\mP}
\providecommand{\mPc}{\bar{\mP}}

\newcommand{\half}{\nicefrac{1}{2}}

\graphicspath{{figures/}}

\begin{document}
\ninept

\title{Diffusion-based Generative Speech Source Separation}

\name{Robin Scheibler$^1$, Youna Ji$^2$, Soo-Whan Chung$^2$, Jaeuk Byun$^2$, Soyeon Choe$^2$, Min-Seok Choi$^2$ %
\thanks{
Code will be shared after the review process.
}}
\address{$^1$LINE Corporation, Tokyo, Japan\\
         $^2$Naver Corporation, South Korea}

%

\maketitle

%

\begin{abstract}
We propose DiffSep, a new single channel source separation method based on score-matching of a stochastic differential equation (SDE).
We craft a tailored continuous time diffusion-mixing process starting from the separated sources and converging to a Gaussian distribution centered on their mixture.
This formulation lets us apply the machinery of score-based generative modelling.
First, we train a neural network to approximate the score function of the marginal probabilities or the diffusion-mixing process.
Then, we use it to solve the reverse time SDE that progressively separates the sources starting from their mixture.
We propose a modified training strategy to handle model mismatch and source permutation ambiguity.
Experiments on the WSJ0\_2mix dataset demonstrate the potential of the method.
Furthermore, the method is also suitable for speech enhancement and shows performance competitive with prior work on the VoiceBank-DEMAND dataset.
\end{abstract}
\begin{keywords}%
source separation, stochastic differential equation, diffusion, score matching
\end{keywords}


\section{Introduction} 
Source separation (SS) refers to a family of techniques that can be used to recover signals of interests from their mixtures.
It has broad applications, but we focus on single channel audio and speech~\cite{makinoAudioSourceSeparation2018}.
This is an underdetermined problem where several signals are recovered from a single mixture.
Early success in this area was made with methods such as non-negative matrix factorization (NMF)~\cite{smaragdis_static_2014}.
Impressive progress was brought by data-driven methods and deep neural networks (DNN).
We mention deep clustering where a DNN is trained to create an embedding space in which the time-frequency bins of the input spectrogram cluster by source~\cite{hersheyDeepClusteringDiscriminative2016}.
Then, one can recover isolated sources by creating masks corresponding to clusters of bins and multiply the spectrogram with it.
These methods then evolved to networks predicting mask values directly~\cite{erdogan_phase-sensitive_2015}.
All these data-driven methods require to solve the fundamental source permutation ambiguity.
Namely, there is no inherent preferred order for the sources.
This is usually handled by permutation invariant training (PIT)~\cite{kolbaekMultitalkerSpeechSeparation2017}.
In PIT, the objective function is computed for all permutations of the sources and the one with minimum value is chosen to compute the gradient.
Recently very strong time-domain baselines such as Conv-TasNet~\cite{luo_conv-tasnet_2019} or dual-path transformer~\cite{chen_dual-path_2020} have been proposed.
We mention also all-attention based~\cite{subakan_attention_2021}, multi-scale networks~\cite{hu_speech_2021},
WaveSplit~\cite{zeghidour_wavesplit_2021},
and the recent state-of-the-art TF-GridNet~\cite{wang_tf-gridnet_2022} back in the time-frequency domain.
For a survey and benchmarks of recent methods see~\cite{lu_espnet-se_2022,wang_tf-gridnet_2022}.
All these recent approaches are trained with a discriminative objective such as the scale-invariant signal-to-distortion ratio (SI-SDR)~\cite{le_roux_sdrhalf-baked_2019}.
Except the traditional methods, such as NMF, few approaches leverage generative models.
One exception adversarially trains speaker-specific generative models and uses them for separation~\cite{subakan_generative_2018}.

Unlike discriminative methods, generative modelling allows to approximate complex data distributions~\cite{bond-taylor_deep_2022}.
Recently, score-based generative modelling (SGM)~\cite{song_generative_2019}, also known as diffusion-based modelling, has made rapid and impressive progress, in particular in the domain of image generation~\cite{rombach_high-resolution_2022}.
SGM defines a forward process that progressively transforms a sample of the target distribution into Gaussian noise.
Given its \textit{score-function}, i.e., the gradient of the log-probability, the target distribution can be sampled by running a reverse process starting from noise.
Crucially, the score-function is (usually) unknown, but can be approximated by a DNN using a simple training strategy~\cite{hyvarinen_estimation_2005}.
Recent approaches result from the combination of score-matching and Langevin sampling~\cite{song_generative_2019,song_improved_2020,jolicoeur-martineau2021adversarial}.
Two frameworks have been proposed to unify various approaches, one based on graphical models~\cite{ho_denoising_2020}, and the other on stochastic differential equations (SDE)~\cite{song_score-based_2022}.
%
SGM has quickly found application in audio and speech processing.
It has been successfully applied in speech synthesis~\cite{kong_diffwave_2020,chen_wavegrad_2020,lee_priorgrad_2022,koizumi_specgrad_2022}, speech restoration~\cite{zhang_restoring_2021,serra_universal_2022}, and bandwidth extension~\cite{han_nu-wave_2022}.
Surprisingly, perhaps, several SGMs have been proposed for speech enhancement, a typically discriminative task~\cite{lu_study_2021,lu_conditional_2022,richter_speech_2022,welker_speech_2022}.
Besides sound, but perhaps closest to this work, separation of images using deep generative priors and Langevin sampling has been demonstrated~\cite{jayaram_source_2020}.
However, to the best of our knowledge, a diffusion-based speech separation method has yet to be proposed.

We fill this gap by proposing what we believe is the first diffusion-based approach for the separation of speech signals.
The proposed algorithm, DiffSep, is based on a carefully designed SDE that starts from the separated signals and converges in average to their mixture.
Then, by solving the corresponding reverse-time SDE, it is possible to recover the individual sources from their mixture.
See \ffref{sde_evolution} for an illustration.
We propose a modified training strategy for the score-matching network that addresses ambiguity in the assignment of the sources at the beginning of the inference process.
We demonstrate the method on the widely used \textit{WSJ0\_2mix} benchmark dataset~\cite{hersheyDeepClusteringDiscriminative2016}.
While it does not outperform, yet, discriminative baselines in terms of separation metrics, it achieves high non-intrusive DNSMOS P.835~\cite{reddy_dnsmos_2022} score, indicating naturalness.
In addition, our approach is also suitable for speech enhancement by considering the noise as just an extra source, and performs comparably to prior work~\cite{richter_speech_2022} on the \textit{VoiceBank-DEMAND} dataset~\cite{valentini-botinhao_speech_2016}.

\section{Background}
\seclabel{background}

\subsection{Notation and Signal Model}
\seclabel{background:notation}

Vectors and matrices are represented by bold lower and upper case letters, respectively.
The norm of vector $\vx$ is written $\| \vx \| = (\vx^\top \vx)^{\half}$.
The $N\times N$ identity matrix is denoted by $\mI_N$.

We represent audio signals with $N$ samples by real valued vectors from $\R^N$.
We introduce a simple mixture model with $K$ sources,
\begin{align}
  \vy = \sum_{k=1}^K \vs_k,
  \elabel{mixing_model}
\end{align}
where $\vs_k \in \R^N$  for all $k$.
We remark that this model can account for environmental noise, reverberation, and other degradation by adding an extra source.
By concatenating all the source signals, we obtain vectors in $\R^{K N}$.
We define in this notation the vector of separated sources and their average value,
\begin{align}
  \vs = [\vs_1^\top, \ldots,\vs_K^\top]^\top,\quad
  \vsbar = K^{-1}[\vy^\top,\ldots,\vy^\top]^\top.
  \elabel{s_vec_def}
\end{align}
We can define the time-invariant mixing operation as a multiplication by $(\mA \otimes \mI_N)$ where $\mA$ is a $K\times K$ matrix, and $\otimes$ is the Kronecker product.
Multiplying a vector $\vv \in \R^{KN}$ by such a matrix amounts to multiplying by $\mA$ all sub-vectors of length $K$ with elements taken from $\vv$ at interval $N$, i.e.,
\begin{align}
  \left( (\mA \otimes \mI_N) \vv \right)_{k N + n} = \sum_{\ell=1}^K A_{k\ell}\, v_{\ell N + n},\quad
  \begin{array}{l}
    k=1,\ldots,K,\\ n=1,\ldots,N.
  \end{array}
\end{align}
To lighten the notation, we overload the regular matrix product as $\mA \vv \triangleq (\mA \otimes \mI_N) \vv$ in the rest of this paper.
Next, we define the projection matrices
\begin{align}
  \mPa = K^{-1} \vone \vone^\top,\quad \mPc = \mI_{K} - \mPa.
  \elabel{proj_mat}
\end{align}
where $\vone$ is the all one vector, of size $K$ here.
The matrix $\mPa$ projects onto the subspace of average values, and $\mPc$ its orthogonal complement.
For example, a compact alternative to \eref{mixing_model} is $\mPa \vs = \vsbar$.

\subsection{SDE for Score-based Generative Modelling}
\seclabel{background:sgm}

Score-based generative modelling (SGM) via SDE is a principled approach to model complex data distributions~\cite{song_score-based_2022}.
First, a diffusion process starting from samples from the target distribution and converging to a Gaussian distribution can be described by the following stochastic differential equation
\begin{align}
  d\vx_t = f(\vx_t, t)\,dt + g(t)\, d\vw_t
  \elabel{sde_forward}
\end{align}
where $\vx_t\,:\, \R^N \to \R^N$ is a vector-valued function of the time $t\in \R$, $d\vx_t$ is its derivative with respect to $t$, and $dt$ an infinitesimal time step.
Functions $f\,:\,\R^N \to \R^N$ and $g\,:\,\R\to\R$ are called the \textit{drift} and \textit{diffusion} coefficients of $\vx_t$.
The term $d\vw_t$ is a standard Wiener process.
For more details on SDEs, see~\cite{sarkka_applied_2019}.
We insist that the time $t$ in the diffusion process is unrelated to the time of the audio signals that will appear later in this paper.

A remarkable property of \eref{sde_forward} is that under some mild conditions, there exists a corresponding \textit{reverse-time SDE},
\begin{align}
  d\vx_t = -[f(\vx_t, t) - g(t)^2 \nabla_{\vx_t} \log p_t(\vx_t) ]\,dt + g(t) \, d\bar{\vw},
  \elabel{sde_backward}
\end{align}
going from $t=T$ to $t=0$~\cite{anderson_reverse-time_1982}.
Here, $d\bar{\vw}$ is a reverse Brownian process, and $p_t(\vx)$ is the marginal distribution of $\vx_t$.

During the forward process, $\vx_0$ being known, it is usually possible to have a closed-form formula for $p_t(\vx)$.
For an affine $f(\vx, t)$, it is Gaussian and its parameters, i.e., mean and covariance matrix, usually tractable~\cite{sarkka_applied_2019}. 
However, in the reverse process, $p_t(\vx)$ is unknown, and thus \eref{sde_backward} cannot be solved directly.
The key idea of SGM~\cite{song_generative_2019,song_improved_2020,song_score-based_2022} is to train a neural network $\vq_{\vtheta}(\vx, t)$ so that $\vq_{\vtheta}(\vx, t)\approx \nabla_{\vx} \log p_t(\vx)$.
Provided that the approximation is good enough, we can generate samples from the distribution of $\vx_0$ by numerically solving \eref{sde_backward}, replacing $\nabla_{\vx}\log p_t(\vx)$ by $\vq_{\vtheta}(\vx, t)$.

\subsection{Related Work in Diffusion-based Speech Processing}

SGM first found widespread adoption for the generation of very high quality synthetic images, e.g.~\cite{rombach_high-resolution_2022}.
Speech synthesis is thus a logical candidate for its adoption~\cite{kong_diffwave_2020,chen_wavegrad_2020}.
In this area, several approaches that shape the noise adaptively for the target speech have been proposed.
One uses the local signal energy~\cite{lee_priorgrad_2022}, while the other introduces correlation in the noise according to the target spectrogram~\cite{koizumi_specgrad_2022}.
Bandwidth extension and restoration also require to generate missing parts of the audio and high quality diffusion-based approaches have been proposed~\cite{han_nu-wave_2022,zhang_restoring_2021,serra_universal_2022}.
Diffuse~\cite{lu_study_2021} and CDiffuse~\cite{lu_conditional_2022} use conditional probabilistic diffusion to enhance a noisy speech signal.
Also for speech enhancement, Richter et al. have proposed another approach~\cite{richter_speech_2022,welker_speech_2022} based on solving the reverse-time SDE corresponding to the forward model
\begin{align}
    d\vx_t = \gamma (\vy - \vx_t)\,dt + g(t)\, d\vw,\quad \vx_0 = \vs.
    \elabel{OUDSDE}
\end{align}
This is the \textit{Ornstein-Uhlenbeck} SDE that, in the forward direction, takes the clean speech $\vs$ progressively towards the noisy one $\vy$.

\section{Diffusion-based Source Separation}
\seclabel{prop:diffeq}

\tikzstyle{every node}=[font=\footnotesize]

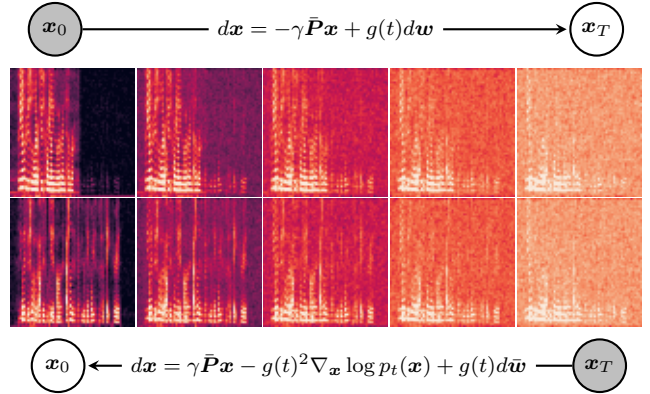
\begin{figure}
    \centering
    \begin{tikzpicture}[->,>=stealth',shorten >=1pt,auto,node distance=3.6cm, thick,main node/.style={circle,draw,minimum size=0.7cm,inner sep=0pt]}] 
    \node[main node,fill=lightgray] (1) {$\vx_0$};
    \node (m) [right of=1] {$d\vx = -\gamma \mPc \vx + g(t) d\vw$};
    \node[main node] (2) [right of=m]  {$\vx_T$};
    
    \draw[-stealth] (1) -- (m) -- (2);
    \end{tikzpicture}
    \smallskip
    
    \includegraphics{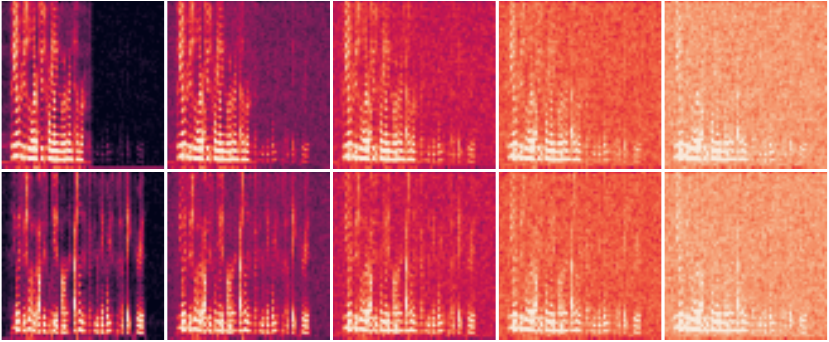}
    \smallskip
    
    \begin{tikzpicture}[->,>=stealth',shorten >=1pt,auto,node distance=3.6cm, thick,main node/.style={circle,draw,minimum size=0.7cm,inner sep=0pt]}] 
    \node[main node] (1) {$\vx_0$};
    \node (m) [right of=1] {$d\vx = \gamma \mPc \vx - g(t)^2 \nabla_{\vx}\log p_t(\vx)+ g(t) d\bar{\vw}$};
    \node[main node,fill=lightgray] (2) [right of=m]  {$\vx_T$};
    
    \draw (2) [-stealth]-- (m) -- (1);
    \end{tikzpicture}
    \caption{Illustration of the diffusion-mixing process. From left to right, the sources start separated and progressively mix as more and more noise is added. The reverse process starts from the mixture and ends with separated sources. The spectrograms were produced by actual separation of the mixture on the right using the proposed method.}
    \flabel{sde_evolution}
\end{figure}

We propose a framework for applying SGM to the source separation problem.
Essentially, we design a forward SDE where both diffusion and mixing across sources happen over time.
Each step of the process can be described as adding an infinitesimal amount of noise and doing an infinitesimal amount of mixing.
This can be formalized as the following SDE
\begin{align}
  d\vx_t & = -\gamma \mPc \vx_t\,dt + g(t) d\vw, \quad \vx_0 = \vs,
  \elabel{mix_sde}
\end{align}
where $\mPc$ is defined in~\eref{proj_mat} and $\vs$ in~\eref{s_vec_def}.
We use the diffusion coefficient of the \textit{Variance Exploding SDE} of Song et al.~\cite{song_score-based_2022},
\begin{align}
    g(t) = \smin \left(\frac{\smax}{\smin}\right)^t \sqrt{2\log\left(\frac{\smax}{\smin}\right)}.
\end{align}
The SDE~\eref{mix_sde} has the interesting property that the mean of the marginal $\vmu_t$ goes from the vector of separated signals at $t=0$ to the mixture vector $\vsbar$ as $t$ grows large.
This is formalized in the following theorem proved in \aref{proof}.
\begin{theorem}
  The marginal distribution of $\vx_t$ is Gaussian with mean and covariance matrix
  \begin{align}
    \vmu_t & = (1 - e^{-\gamma t}) \vsbar + e^{-\gamma t} \vs \elabel{xt_mean} \\
    \mSigma_t & = \lambda_1(t) \mPa + \lambda_2(t) \mPc \elabel{xt_cov}
  \end{align}
  where if we let $\xi_1 = 0$, $\xi_2 = \gamma$, and $\rho=\frac{\smax}{\smin}$, then,
  \begin{align}
    \lambda_k(t) & = \frac{\smin^2 \left( \rho^{2t} - e^{-2\xi_k t} \right) \log\rho}{\xi_k + \log\rho}.
  \end{align}
  \label{THM:MARGINAL}
\end{theorem}
\ffref{sde_evolution} shows an example of the process on a real mixture.
We now have an explicit expression for a sample $\vx_t$,
\begin{align}
  \vx_t = \vmu_t + \mL_t \vz,\quad \vz \sim \calN(\vzero, \mI_{KN}),
  \elabel{x_t}
\end{align}
where $\mL_t = \lambda_1(t)^{\half} \mPa + \lambda_2(t)^{\half} \mPc$.
\ffref{sde_evo} shows the evolution of some of the parameters of the distribution of $\vx_t$ over time.
We can make the difference of $\vmu_T$ and $\vsbar$ arbitrarily small by tuning the value of $\gamma$.
In addition, we observe that the correlation of the noise added onto both sources increases over time due to the mixing process.

Previous work on speech enhancement has successfully applied the diffusion process in the short-time Fourier transform (STFT) domain with a non-linear transform~\cite{welker_speech_2022,richter_speech_2022}.
The approach we propose could be equally applied in the time-domain or in any linearly transformed domain such as the STFT.
However, since the process of \eref{mix_sde} models the linear mixing of the sources, we cannot apply a non-linear transformation.
Instead, we perform the diffusion process in the time-domain, but operate our network in the non-linear STFT domain of~\cite{welker_speech_2022}, as described in \sref{experiments:network}.

\begin{figure}
  \centering
  \includegraphics[width=\linewidth]{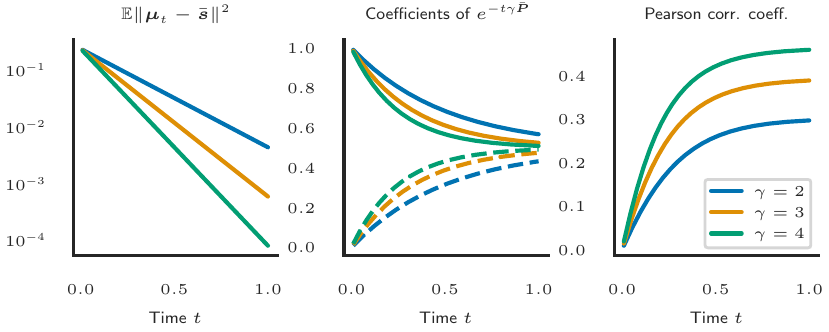}
  \caption{Evolution of some of the parameters of the SDE \eref{mix_sde} over time.}
  \flabel{sde_evo}
\end{figure}

\subsection{Inference}

At inference time, the separation is done by solving \eref{sde_backward} with the help of a score-matching network $\vq_{\vtheta}(\vx, t, \vy) \approx \nabla_{\vx} \log p_t(\vx)$.
The initial value of the process is sampled from the distribution,
\begin{align}
  \bar{\vx}_T \sim \calN(\vsbar, \mSigma_T \mI_{KN}).
  \elabel{prior_sampling}
\end{align}
Then, we adopt the predictor-corrector approach described in~\cite{song_score-based_2022} to solve \eref{sde_backward}.
The prediction step is done by \textit{reverse diffuse sampling}.
The correction is done by annealed Langevin sampling~\cite{jolicoeur-martineau2021adversarial}.

\subsection{Permutation and Mismatch-aware Training Procedure}
\seclabel{prop:score_matching}

To train the score network $\vq_{\vtheta}(\vx, t, \vy)$, we propose a slightly modified score-matching procedure.
But, first, we follow the usual procedure.
Thanks to Theorem~\ref{THM:MARGINAL}, the score function has a closed form expression.
The computation is similar to that in~\cite{richter_speech_2022}, and we obtain
\begin{align}
    \nabla_{\vx_t}\log p(\vx_t) = -\mSigma_t^{-1} (\vx_t - \vmu_t) = -\mL_t^{-1} \vz.
\end{align}
The training loss is 
\begin{align}
  \calL & = \mathbb{E}_{\vx_0,\vz,t} \left\| \vq_\theta(\vx_t,t,\vy) + \mL_t^{-1} \vz\right\|_{\mSigma_t}^2 \\
  & = \mathbb{E}_{\vx_0,\vz,t} \left\| \mL_t \vq_\theta(\vx_t,t,\vy) + \vz\right\|^2,
  \elabel{loss_score_matching}
\end{align}
where $\vz\sim\calN(\vzero, \mI_{KN})$, $t\sim\calU(t_\epsilon, T)$, and $\vx_0$ is chosen at random from the dataset.
We use the norm induced by $\mSigma_t$ which is equivalent to the weighting scheme proposed by Song et al.~\cite{song_generative_2019}.
This scheme makes the order of magnitude of the cost independent of $\mSigma_t$ and lets us avoid the computation of the inverse of $\mL_t$.

However, when training only with the procedure just described, we found poor performance at inference time.
We identified two problems.
First, there is a discrepancy between $\mathbb{E}[\bar{\vx}_T] = \vsbar$ from \eref{prior_sampling} and $\vmu_T$.
While $\vmu_T$ contains slightly different ratios of the sources in its components, $\vsbar$ does not.
Second, the network needs to decide in which order to output the sources.
This is usually solved with a PIT objective~\cite{kolbaekMultitalkerSpeechSeparation2017}.
To solve these challenges, we propose to train the network in a way that includes the model mismatch.

For each sample, with probability $1 - p_T$, $p_T\in[0, 1]$, we apply the regular score-matching procedure minimizing~\eref{loss_score_matching}.
With probability $p_T$, we set $t=T$ and minimize the following alternative loss,
\begin{align}
  \calL_{T} =
  \mathbb{E}_{\vx_0,\vz} \min_{\pi \in \calP} \left\| \mL_T \vq_\theta(\bar{\vx}_T,T,\vy)
  + \vz + \mL_T^{-1} (\vsbar - \vmu_T(\pi)) \right\|^2,
  \nonumber
\end{align}
where $\bar{\vx}_T$ is sampled from \eref{prior_sampling}, $\calP$ is the set of permutations of the sources, and $\vmu_T(\pi)$ is $\vmu_T$ computed for the permutation $\pi$.
The rational for this objective is that the score network will learn to remove both the noise and the model mismatch.
We note that for speech enhancement, there is no ambiguity in the order of sources and the minimization over permutations can be removed.

\section{Experiments}

\subsection{Datasets}

We use the \textit{WSJ0\_2mix} dataset~\cite{hersheyDeepClusteringDiscriminative2016} for the speech separation experiment.
It contains 20000 (\SI{30}{\hour}), 5000 (\SI{10}{\hour}), and 3000 (\SI{5}{\hour}) two speakers mixtures whose individual speech samples were taken from the WSJ0 dataset.
The utterances are fully overlapped and speakers do not overlap between training and test datasets.
The relative signal-to-noise ratio (SNR) of the mixtures is from \SIrange{-5}{5}{\decibel}.
The sampling frequency is \SI{8}{\kilo\hertz}.
In addition, we use the clean test set of Libri2Mix~\cite{cosentino_librimix_2020} to test mismatched conditions.

For the speech enhancement task, we use the \textit{VoiceBank-DEMAND} dataset~\cite{valentini-botinhao_speech_2016}.
It is a collection of clean speech samples of 30 speakers from the VoiceBank corpus degraded by noise, recorded as well as artificial.
The dataset is split into train (\SI{8.6}{\hour}), validation (\SI{0.7}{\hour}), and test (\SI{0.6}{\hour}).
The SNR of the mixtures ranges from \SIrange{0}{15}{\decibel} for training and validation, and from \SIrange{2.5}{17.5}{\decibel} for testing.
The sampling rate of the mixtures is \SI{16}{\kilo\hertz}.

\subsection{Model}
\seclabel{experiments:network}

We use the noise conditioned score-matching network (NCSN++) of Song et al.~\cite{song_score-based_2022}.
An STFT and its inverse layer sandwich the network and, in addition, we apply the non-linear $c(x)$ transform proposed in~\cite{welker_speech_2022} after the STFT, and its inverse before the iSTFT,
\begin{align}
  c(x) = \beta^{-1} |x|^\alpha e^{j \angle x},\quad c^{-1}(x) = \beta |x|^{\nicefrac{1}{\alpha}} e^{j \angle x}.
  \elabel{transform}
\end{align}
The real and imaginary parts are concatenated.
The network has $K+1$ inputs, one for each of the target sources plus one for the mixture on which we condition, and $K$ outputs.

\subsection{Hyperparameters for Inference and Training}

For inference, we use 30 prediction steps. For each one, we do one correction step with step size $r=0.5$.
The networks are trained using Pytorch with the Adam optimizer.
We use exponential averaging of the network weights as in~\cite{richter_speech_2022}.

\textbf{Separation}
We use $\gamma=2$, $\sigma_{\min}=0.05$, $\sigma_{\max}=0.5$.
For the transformation~\eref{transform}, we use $\alpha=0.5$ and $\beta=0.15$.
The probability of selecting a sample for modified training is $p_{T}=0.1$.
The effective batch size is 48 and the learning rate $0.0005$.
We trained for 1000 epochs and we chose the one with largest SI-SDR on a sentinel validation batch.

\textbf{Enhancement}
The SDE parameters are $\gamma=2$, $\sigma_{\min}=0.05$, $\sigma_{\max}=0.5$.
The transformation~\eref{transform} uses $\alpha=0.5$ and $\beta=0.15$.
We found it beneficial to use PriorGrad~\cite{lee_priorgrad_2022} for the noise generation. The variance of the process noise for a given sample is chosen as the average power of the 500 closest samples in the mixture.
In the enhancement setting, there is no permutation ambiguity so we always order the sources with speech first and noise second.
We use $p_T=0.03$.
The effective batch size was 16 and the learning rate 0.0001.
The network was trained for around 160 epochs.

\begin{figure}
    \centering
    \includegraphics{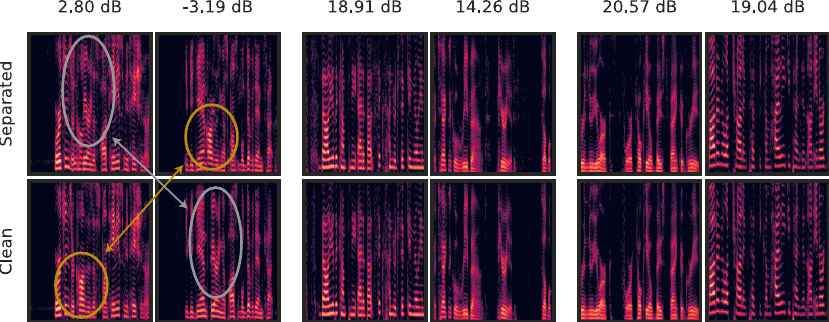}
    \caption{Spectrograms of the clean sources (bottom row) and the separated speech obtained by applying the proposed method to their mixture (top row). The three samples, from left to right, were chosen randomly from the bottom \SI{10}{\percent}, mid \SI{80}{\percent}, and top \SI{10}{\percent} samples sorted by SI-SDR.
    The SI-SDR of the samples is indicated at the top.
    In the low quality sample, the spectrogram looks good, but we can see block permutations of the sources occurring.}
    \flabel{samples}
\end{figure}

\subsection{Results}

We measure the performance in terms of SI-SDR~\cite{le_roux_sdrhalf-baked_2019}, perceptual evaluation of speech quality (PESQ)~\cite{rix_perceptual_2001}, extended short-time objective intelligibility (ESTOI)~\cite{jensen_algorithm_2016},  and the overall (OVRL) metric of the deep noise challenge mean opinion score (DNSMOS) P.835~\cite{reddy_dnsmos_2022}.

\begin{table}
\footnotesize
  \centering
  \caption{Separation results on WSJ0\_2mix and Libri2Mix (clean) datasets.}
  \begin{tabular}{@{}llrrrr@{}}
    \toprule
    Dataset & Model & SI-SDR & PESQ & ESTOI & OVRL \\
    \midrule
    WSJ0\_2mix & Conv-TasNet~\cite{luo_conv-tasnet_2019} & 16.0 & 3.29 & 0.91 & 3.21 \\
    {\scriptsize (matched)} & DiffSep {\scriptsize(proposed)} &  14.3  & 3.14 & 0.90 & 3.29 \\  
    \midrule
    Libri2Mix & Conv-TasNet~\cite{luo_conv-tasnet_2019} & 10.3  & 2.60 & 0.80 & 2.95 \\ 
    {\scriptsize (mismatched)} & DiffSep {\scriptsize(proposed)} & 9.6 & 2.58 & 0.78 & 3.12 \\ 
    \bottomrule
  \end{tabular}
  \tlabel{sep_results}
\end{table}

\begin{table}
\footnotesize
  \centering
  \caption{Enhancement results on the VoiceBank-DEMAND dataset.}
  \begin{tabular}{@{}lrrrr@{}}
    \toprule
    Model & SI-SDR & PESQ & ESTOI & OVRL \\
    \midrule
    Conv-TasNet~\cite{luo_conv-tasnet_2019}  & 18.3 & 2.88 & 0.86 & 3.20 \\
    \midrule
    CDiffuse$^{\dagger}$~\cite{lu_conditional_2022}      & 12.6 & 2.46 & 0.79 & --- \\
    SGMSE+$^{\dagger}$~\cite{richter_speech_2022}        & 17.3 & 2.93 & 0.87 & --- \\
    DiffSep {\scriptsize(proposed)}                                 & 17.5 & 2.56 & 0.84 & 3.09 \\
    \bottomrule
    \multicolumn{5}{@{}l@{}}{$^{\dagger}$~results reported in~\cite{richter_speech_2022}.} \\
  \end{tabular}
  \tlabel{enh_results}
\end{table}

\textbf{Separation}
\tref{sep_results} shows the results of the speech separation experiments.
We also train a baseline Conv-TasNet~\cite{luo_conv-tasnet_2019} network.
On the intrusive metrics, SI-SDR, PESQ, and STOI, the proposed method scores lower than the baseline.
However, it scores higher on the non-intrusive OVRL metric.
Informal listening tests further reveal that low separation metrics are often due to very natural sounding permutations of the speakers.
This problem seems to happen for similar sounding speakers.
Example spectrograms of poor, fair, and high quality separation results are shown in \ffref{samples}.


It was reported that diffusion-based enhancement performs well in mismatched condition~\cite{richter_speech_2022}.
We thus evaluated the model trained on WSJ0\_2mix on the Libri2Mix~\cite{cosentino_librimix_2020} clean test set.
The proposed method still scores lower than Conv-TasNet for SI-SDR, PESQ, and STOI, but the metric degradation is smaller, e.g. \SI{1}{\decibel} less reduction of SI-SDR.
On the other hand, the proposed method still enjoys fairly high OVRL score, indicating good perceptual quality.

\textbf{Enhancement}
\tref{enh_results} shows the results of the evaluation of the noisy speech enhancement task.
The proposed method performs very similarly to other diffusion methods CDiffuse~\cite{lu_conditional_2022} and SGMSE+~\cite{richter_speech_2022}.
We obtain an SI-SDR higher by \SI{0.2}{\decibel}, but PESQ value closer to that of CDiffuse.
ESTOI is somewhere in the middle.
The OVRL metric suggests good perceptual quality, but was not provided for the other methods.
In terms of SI-SDR particularly, a gap remains with discriminative methods such as Conv-TasNet.

\section{Conclusion}

We proposed DiffSep, the first speech source separation method based on diffusion process.
We successfully separated speech sources, and demonstrated that our approach extends to noisy speech enhancement.
For separation, there was a substantial gap in performance with state-of-the-art discriminative methods.
We believe this may be due to both outputs being from the same distribution, thus confusing the generative model.
We have done initial work to alleviate the permutation ambiguity, using a modified training strategy for the score-matching network, but further work is likely necessary to robustify the inference procedure against permutation mistakes.
Bringing in proven architectures from the separation literature is an interesting area of future work.
We have also not used yet any of the theoretical properties of the distribution of mixtures, which could be elegantly integrated in the score network architecture, as in~\cite{jayaram_source_2020}.

\vfill

\section{Appendix: Proof of Theorem~\ref{thm:marginal}}
\seclabel{proof}

Because the stochastic differential equation is linear, $\vx_t$ follows a Gaussian distribution~\cite{anderson_reverse-time_1982}.
Following S\"{a}rkk\"{a} and Solin~\cite{sarkka_applied_2019}, eq. (5.50) and (5.53), the mean and covariance matrices of $\vx_t$ are the solutions of
\begin{align}
  d\vmu_t = -\mPc \vmu_t,\quad 
  d\mSigma_t = -\mAhat \mSigma_t - \mSigma_t \mAhat + g(t)^2 \mI,
\end{align}
respectively, with the initial conditions $\vmu_0 = \vx_0$ and $\mSigma_0 = \vzero$.
The solution for $d\vmu_t$ is given by the the matrix exponential operator,
\begin{align}
  \vmu_t = e^{-\gamma t \mPc} \vx_0 = exp(0) \mPa \vx_0 + exp(-\gamma t) (\mI - \mPa) \vx_0.
\end{align}
Substituting $\vx_0 = \vs$, $\mP \vx_0 = \vsbar$, and rearranging terms yields \eref{xt_mean}.
For $\mSigma_t$, we work in the eigenbasis of $\gamma \mPc$ and solve differential equations with respect to its distinct eigenvalues, $0$ and $\gamma$. \hfill \ensuremath{\square}

\clearpage

\section{References}

\begingroup
\setlength\bibitemsep{0pt}
\printbibliography[heading=none]

@STRING{P_ICASSP       = "ICASSP"}

@STRING{P_INTERSPEECH  = "INTERSPEECH"}

@STRING{P_MMSP         = "MMSP"}

@STRING{P_APSIPA       = "APSIPA"}

@STRING{P_CVPR         = "CVPR"}

@STRING{P_ICML         = "{ICML}"}

@STRING{P_ICLR         = "{ICLR}"}

@book{makinoAudioSourceSeparation2018,
	address = {Cham},
	series = {Signals and {Communication} {Technology}},
	title = {Audio {Source} {Separation}},
	publisher = {Springer International Publishing},
	editor = {Makino, Shoji},
	month = jan,
	year = {2018},
}

@inproceedings{hersheyDeepClusteringDiscriminative2016,
	%address = {Shanghai, CN},
	title = {Deep clustering: {Discriminative} embeddings for segmentation and separation},
	booktitle = P_ICASSP,
	author = {Hershey, John R and Chen, Zhuo and Le Roux, Jonathan and Watanabe, Shinji},
	month = mar,
	year = {2016},
	pages = {31--35}
}

@inproceedings{lu_espnet-se_2022,
	title = {{ESPnet}-{SE}++: {Speech} Enhancement for Robust Speech Recognition, Translation, and Understanding},
	booktitle = P_INTERSPEECH,
	author = {Lu, Yen-Ju and Chang, Xuankai and Li, Chenda and Zhang, Wangyou and Cornell, Samuele and Ni, Zhaoheng and Masuyama, Yoshiki and Yan, Brian and Scheibler, Robin and Wang, Zhong-Qiu and Tsao, Yu and Qian, Yanmin and Watanabe, Shinji},
	month = sep,
	year = {2022},
	pages = {5458--5462}
}

@inproceedings{chen_dual-path_2020,
	title = {Dual-Path Transformer Network: Direct Context-Aware Modeling for End-to-End Monaural Speech Separation},
	booktitle = P_INTERSPEECH,
	author = {Chen, Jingjing and Mao, Qirong and Liu, Dong},
	month = oct,
	year = {2020},
	pages = {2642--2646},
}

@inproceedings{erdogan_phase-sensitive_2015,
	address = {Brisbane, AUD},
	title = {Phase-sensitive and recognition-boosted speech separation using deep recurrent neural networks},
	booktitle = P_MMSP,
	author = {Erdogan, H and Hershey, J R and Watanabe, Shinji and Le Roux, Jonathan},
	month = apr,
	year = {2015},
	pages = {708--712},
}

@article{kolbaekMultitalkerSpeechSeparation2017,
	title = {Multitalker Speech Separation With Utterance-Level Permutation Invariant Training of Deep Recurrent Neural Networks},
	volume = {25},
	number = {10},
	journal = {IEEE/ACM Trans. Audio Speech Lang. Process.},
	author = {Kolbaek, Morten and Yu, Dong and Tan, Zheng-Hua and Jensen, Jesper},
	month = aug,
	year = {2017},
	pages = {1901--1913}
}

@inproceedings{song_improved_2020,
	title = {Improved Techniques for Training Score-Based Generative Models},
  booktitle = {Adv. neural inf. process. syst.},
  volume = {33},
  pages = {12438--12448},
	author = {Song, Yang and Ermon, Stefano},
	month = jun,
	year = {2020}
}

@inproceedings{ho_denoising_2020,
  title = {Denoising {Diffusion} {Probabilistic} {Models}},
  volume = {33},
  booktitle = {Adv. Neural Inf. Process. Syst.},
  publisher = {Curran Associates, Inc.},
  author = {Ho, Jonathan and Jain, Ajay and Abbeel, Pieter},
  year = {2020},
  pages = {6840--6851}
}

@misc{richter_speech_2022,
	title = {Speech Enhancement and Dereverberation with Diffusion-based Generative Models},
	author = {Richter, Julius and Welker, Simon and Lemercier, Jean-Marie and Lay, Bunlong and Gerkmann, Timo},
	month = aug,
	year = {2022},
  note={arXiv:2208.05830}
}

@inproceedings{lu_conditional_2022,
	title = {Conditional Diffusion Probabilistic Model for Speech Enhancement},
	booktitle = P_ICASSP,
	author = {Lu, Yen-Ju and Wang, Zhong-Qiu and Watanabe, Shinji and Richard, Alexander and Yu, Cheng and Tsao, Yu},
	month = may,
	%address = {Singapore, SG},
	year = {2022},
	pages = {7402--7406},
}

@inproceedings{lu_study_2021,
	title = {A Study on Speech Enhancement Based on Diffusion Probabilistic Model},
	booktitle = P_APSIPA,
	%address = {Tokyo, JP},
	author = {Lu, Yen-Ju and Tsao, Yu and Watanabe, Shinji},
	month = dec,
	year = {2021},
	pages = {659--666}
}

@inproceedings{welker_speech_2022,
	title = {Speech Enhancement with Score-Based Generative Models in the Complex {STFT} Domain},
	booktitle = P_INTERSPEECH,
	%address = {Incheon, KR},
	author = {Welker, Simon and Richter, Julius and Gerkmann, Timo},
	month = sep,
	year = {2022},
	pages = {2928--2932},
}

@inproceedings{zhang_restoring_2021,
	title = {Restoring degraded speech via a modified diffusion model},
	author = {Zhang, Jianwei and Jayasuriya, Suren and Berisha, Visar},
	booktitle = P_INTERSPEECH,
	%address = {Brno, CZ},
	month = apr,
	year = {2021},
	pages = {221--225}
}

@article{anderson_reverse-time_1982,
	title = {Reverse-time diffusion equation models},
	volume = {12},
	number = {3},
	journal = {Stochastic Processes and their Applications},
	author = {Anderson, Brian D. O.},
	month = may,
	year = {1982},
	pages = {313--326},
}

@inproceedings{song_score-based_2022,
	title = {Score-Based Generative Modeling through Stochastic Differential Equations},
	author = {Song, Yang and Sohl-Dickstein, Jascha and Kingma, Diederik P. and Kumar, Abhishek and Ermon, Stefano and Poole, Ben},
	booktitle = P_ICLR,
	month = may,
	year = {2021},
}

@inproceedings{song_generative_2019,
	title = {Generative Modeling by Estimating Gradients of the Data Distribution},
	volume = {32},
	booktitle = {Adv. Neural Inf. Process. Syst.},
	author = {Song, Yang and Ermon, Stefano},
	year = {2019},
}

@inproceedings{jolicoeur-martineau2021adversarial,
  title={Adversarial score matching and improved sampling for image generation},
  author={Alexia Jolicoeur-Martineau and R{\'e}mi Pich{\'e}-Taillefer and Ioannis Mitliagkas and Remi Tachet des Combes},
  booktitle=P_ICLR,
  year={2021},
  month=may
}

@book{sarkka_applied_2019,
	edition = {1},
	title = {Applied {Stochastic} {Differential} {Equations}},
	publisher = {Cambridge University Press},
	author = {S\"{a}rkk\"{a}, Simo and Solin, Arno},
	month = apr,
	year = {2019},
}

@inproceedings{kong_diffwave_2020,
	title = {{DiffWave}: A Versatile Diffusion Model for Audio Synthesis},
  booktitle = P_ICLR,
  year = 2021,
	author = {Kong, Zhifeng and Ping, Wei and Huang, Jiaji and Zhao, Kexin and Catanzaro, Bryan},
	month = may
}

@inproceedings{chen_wavegrad_2020,
	title = {{WaveGrad}: Estimating Gradients for Waveform Generation},
	author = {Chen, Nanxin and Zhang, Yu and Zen, Heiga and Weiss, Ron J and Norouzi, Mohammad and Chan, William},
	month = may,
	year = {2021},
  booktitle=P_ICLR

}

@inproceedings{lee_priorgrad_2022,
	title = {{PriorGrad}: Improving Conditional Denoising Diffusion Models with Data-Dependent Adaptive Prior},
	booktitle = P_ICLR,
	author = {Lee, Sang-gil and Kim, Heeseung and Shin, Chaehun and Tan, Xu and Liu, Chang and Meng, Qi and Qin, Tao and Chen, Wei and Yoon, Sungroh and Liu, Tie-Yan},
	month = apr,
	year = {2022}
}

@inproceedings{koizumi_specgrad_2022,
	title = {{SpecGrad}: Diffusion Probabilistic Model based Neural Vocoder with Adaptive Noise Spectral Shaping},
	booktitle = P_INTERSPEECH,
	author = {Koizumi, Yuma and Zen, Heiga and Yatabe, Kohei and Chen, Nanxin and Bacchiani, Michiel},
	month = sep,
	year = {2022},
	pages = {803--807},
  %address = {Incheon, KR}
}

@misc{serra_universal_2022,
	title = {Universal Speech Enhancement with Score-based Diffusion},
	author = {Serr\`{a}, Joan and Pascual, Santiago and Pons, Jordi and Araz, R. Oguz and Scaini, Davide},
	month = jun,
	year = {2022},
  note = {Arxiv:2206.03065}
}

@inproceedings{valentini-botinhao_speech_2016,
	%address = {San Francisco, USA},
	title = {Speech Enhancement for a Noise-Robust Text-to-Speech Synthesis System Using Deep Recurrent Neural Networks},
	booktitle = P_INTERSPEECH,
	author = {Valentini-Botinhao, Cassia and Wang, Xin and Takaki, Shinji and Yamagishi, Junichi},
	month = sep,
	year = {2016},
	pages = {352--356}
}

@article{smaragdis_static_2014,
	title = {Static and Dynamic Source Separation Using Nonnegative Factorizations: A unified view},
	volume = {31},
	number = {3},
	journal = {IEEE Signal Process. Mag.},
	author = {Smaragdis, Paris and Fevotte, Cedric and Mysore, Gautham J and Mohammadiha, Nasser and Hoffman, Matthew},
	month = apr,
	year = {2014},
	pages = {66--75}
}

@inproceedings{subakan_generative_2018,
	title = {Generative Adversarial Source Separation},
	booktitle = P_ICASSP,
	author = {Subakan, Y Cem and Smaragdis, Paris},
	year = {2018},
	pages = {26--30},
  %address = {Calgary, CA},
  month = apr
}

@inproceedings{jayaram_source_2020,
	title = {Source Separation with Deep Generative Priors},
	booktitle = P_ICML,
	publisher = {PMLR},
	author = {Jayaram, Vivek and Thickstun, John},
	month = nov,
	year = {2020},
	pages = {4724--4735}
}

@article{luo_conv-tasnet_2019,
	title = {Conv-{TasNet}: Surpassing Ideal Time–Frequency Magnitude Masking for Speech Separation},
	volume = {27},
	number = {8},
	journal = {IEEE/ACM Trans. Audio Speech Lang. Process.},
	author = {Luo, Yi and Mesgarani, Nima},
	month = aug,
	year = {2019},
	pages = {1256--1266}
}

@inproceedings{le_roux_sdrhalf-baked_2019,
  %address = {Brighton, UK},
  title = {{SDR}–half-baked or well done?},
  booktitle = P_ICASSP,
  author = {Le Roux, J and Wisdom, S and Erdogan, Hakan and Hershey, John R},
  month = may,
  year = {2019},
  pages = {626--630}
}

@article{bond-taylor_deep_2022,
  title = {Deep Generative Modelling: A Comparative Review of {VAEs}, {GANs}, Normalizing Flows, Energy-Based and Autoregressive Models},
  volume = {44},
  number = {11},
  journal = {IEEE Trans. Pattern Anal. Mach. Intell.},
  author = {Bond-Taylor, Sam and Leach, Adam and Long, Yang and Willcocks, Chris G.},
  month = nov,
  year = {2022},
  pages = {7327--7347}
}

@inproceedings{rombach_high-resolution_2022,
	title = {High-Resolution Image Synthesis With Latent Diffusion Models},
  booktitle = P_CVPR,
	author = {Rombach, Robin and Blattmann, Andreas and Lorenz, Dominik and Esser, Patrick and Ommer, Björn},
	year = {2022},
  month = jun,
  %address = {New Orleans, LA, USA},
	pages = {10684--10695},
}

@article{hyvarinen_estimation_2005,
	title = {Estimation of Non-Normalized Statistical Models by Score Matching},
	volume = {6},
	journal = {J. Mach. Learn. Res.},
	author = {Hyv\"{a}rinen, Aapo},
	month = apr,
	year = {2005},
	pages = {695--709},
}

@inproceedings{reddy_dnsmos_2022,
	title = {Dnsmos {P}.835: A Non-Intrusive Perceptual Objective Speech Quality Metric to Evaluate Noise Suppressors},
	booktitle = P_ICASSP,
	author = {Reddy, Chandan K A and Gopal, Vishak and Cutler, Ross},
	month = may,
	year = {2022},
  %address = {Singapore, SG},
	pages = {886--890},
}

@inproceedings{rix_perceptual_2001,
	title = {Perceptual evaluation of speech quality ({PESQ}) --- a new method for speech quality assessment of telephone networks and codecs},
	author = {Rix, A W and Beerends, J G and Hollier, M P and Hekstra, A P},
  booktitle = P_ICASSP,
	month = jan,
	year = {2001},
	pages = {749--752},
}

@article{jensen_algorithm_2016,
	title = {An Algorithm for Predicting the Intelligibility of Speech Masked by Modulated Noise Maskers},
	volume = {24},
	number = {11},
	journal = {IEEE/ACM Trans. Audio Speech Lang. Process.},
	author = {Jensen, Jesper and Taal, Cees H.},
	month = nov,
	year = {2016},
	pages = {2009--2022}
}

@inproceedings{han_nu-wave_2022,
	title = {{NU}-{Wave} 2: A General Neural Audio Upsampling Model for Various Sampling Rates},
	booktitle = P_INTERSPEECH,
	author = {Han, Seungu and Lee, Junhyeok},
	month = sep,
	year = {2022},
	pages = {4401--4405},
    address= {Incheon, KR}
}

@misc{cosentino_librimix_2020,
	title = {{LibriMix}: An Open-Source Dataset for Generalizable Speech Separation},
	author = {Cosentino, Joris and Pariente, Manuel and Cornell, Samuele and Deleforge, Antoine and Vincent, Emmanuel},
	month = may,
	year = {2020},
	note = {arXiv:2005.11262 [eess]}
}

@misc{wang_tf-gridnet_2022,
	title = {{TF}-{GridNet}: Making Time-Frequency Domain Models Great Again for Monaural Speaker Separation},
	author = {Wang, Zhong-Qiu and Cornell, Samuele and Choi, Shukjae and Lee, Younglo and Kim, Byeong-Yeol and Watanabe, Shinji},
	month = sep,
	year = {2022},
	note = {arXiv:2209.03952},
}

@inproceedings{hu_speech_2021,
	title = {Speech Separation Using an Asynchronous Fully Recurrent Convolutional Neural Network},
	volume = {34},
	booktitle = {Adv. Neural Inf. Process. Syst.},
	author = {Hu, Xiaolin and Li, Kai and Zhang, Weiyi and Luo, Yi and Lemercier, Jean-Marie and Gerkmann, Timo},
	year = {2021},
	pages = {22509--22522},
}

@inproceedings{subakan_attention_2021,
	title = {Attention Is All You Need In Speech Separation},
	booktitle = P_ICASSP,
	author = {Subakan, Cem and Ravanelli, Mirco and Cornell, Samuele and Bronzi, Mirko and Zhong, Jianyuan},
	month = jun,
	year = {2021},
	pages = {21--25},
	% address = {Toronto, CA}
}

@article{zeghidour_wavesplit_2021,
	title = {Wavesplit: End-to-End Speech Separation by Speaker Clustering},
	volume = {29},
	journal = {IEEE/ACM Trans. Audio Speech Lang. Process.},
	author = {Zeghidour, Neil and Grangier, David},
	year = {2021},
	pages = {2840--2849},
}
\endgroup

\end{document}